\documentclass[12pt,twoside]{article}
\usepackage{a4wide}
\usepackage[T1]{fontenc}
\usepackage{amsmath,amssymb,amsfonts}
\usepackage{times}

\def\IM{\mathop{\Im{\rm m}}\nolimits}
\begin{document}
\title{The Froissart bound  for inelastic cross-sections}
\author{\textbf{Andr\'e Martin}\\
{\small CERN, Theory Division}\\[-3pt]
{\small CH1211 Geneve 23, Switzerland}}
\date{\small\today}
\maketitle
\begin{abstract}
\noindent
We prove that while the total cross{}-section is bounded by
$(\pi/m_\pi^2) \ln^2 s$, where $s$ is the square of the c.m.\ energy and $m_\pi$
the mass of the pion, the
total inelastic cross{}-section is bounded by $(1/4)(\pi/m_\pi^2) \ln^2 s$, which is 4 times smaller.  We discuss the implications of this result on the total
cross{}-section itself.
\end{abstract}

The Froissart bound \cite{r1}, proved later from local massive field
theory and unitarity \cite{r2}, is generally written as.
\begin{equation}\label{eq:FB}
\sigma_T<\frac{\pi}{m_\pi^2} \,\ln^2 s~,
\end{equation}
where $s$ is the square of the c.m.\ energy and $m_\pi$ the pion mass.

The constant in front of $\ln^2 s$ was obtained by
L.~Lukaszuk and myself \cite{r3}. Many of my friends, especially Peter Landshoff, complained that this constant is much too large. It is true that some fits of the proton{}-proton and proton{}-antiproton cross{}-sections \cite{r3} indicate
the possible presence of a ln s square term with, however a much smaller coefficient, about 500 times smaller. Joachim Kupsch, Shasanka Roy,
David Atkinson, Porter Johnson, and myself  are planning to try to improve
this constant by taking into account analyticity, unitarity, including
\emph{elastic} unitarity in the elastic region. To date, there is no example of
amplitude satisfying these requirements. Atkinson \cite{r5} has produced amplitudes satisfying all requirements but where $\sigma_T \propto \ln^{-3}s$.

If we want to undertake such a programme a preliminary
requirement is to start on a well defined basis. It has been recognized long ago that
the Froissart bound is non local by Common \cite{r6} and Yndurain \cite{r7}. See also
\cite{r8,r9}.  Namely, one has in fact:
\begin{equation}\label{eq:FBA}
s^N\,\int_s^{s+1/s^{N}}\sigma_T(s')\,\mathrm{d} s'< C_N \ln^2 s~.
\end{equation}
The constant $C_N$ , however, depends on $N$. The narrower is the interval, the larger is
$C_N$. This comes from the fact that the basic ingredient of the Froissart
bound is the convergence of the integral
\begin{equation}\label{eq:int1}
\int_{s_0}^\infty\frac{A_s(s,t)}{s^3}\,\mathrm{d} s <\infty~,
\end{equation}
for $0< t\le 4 m_\pi^2$ (sometimes only $0< t< 4 m_\pi^2$,
strictly!), where $A_s$ is the absorptive part of the scattering amplitude, and  $t$ the square of the momentum transfer
\begin{equation}\label{eq:def-ts}
t=2k^2(\cos\theta-1)~,\qquad s=\left(\sqrt{M_A^2+k^2}+\sqrt{M_B^2+k^2}\right)^2~.
\end{equation}

We have explained in \cite{r9} and will show elsewhere \cite{r10} that, if one
wants to preserve the value of the constant in (\ref{eq:FB}),  the average should be
taken on a large interval, for instance:
\begin{equation}\label{eq:FBB}
\overline{\sigma_T}(s)=\frac{1}{s}\,\int_s^{2s}\sigma_T(s')\,\mathrm{d}s'
<\frac{\pi}{m_\pi^2}\,\ln^2 s+ A \ln s + B~,
\end{equation}

where  $A$ and $B$ are determined by low energy parameters in the $t$ channel.

Here we want to report something different and seeming
naively obvious, namely that for the inelastic cross section $\sigma_I$,
\begin{equation}\label{eq:inegi}
\sigma_I<\frac{\pi}{4 m_\pi^2} \,\ln^2 s~,
\end{equation}
The bound is 4 times smaller than the one on the total
cross{}-section.

If there was a strictly sharp cut{}-off in the partial
wave distribution, this would indeed be obvious, because if the scattering amplitude $F(s,t)$ is given by
\begin{equation}\label{eq:pw}
F(s,t)=\frac{\sqrt{s}}{2 k} \sum_\ell (2\ell+1)\,f_\ell(s)\,P_\ell\left(1+\frac{t}{2 k^2}\right)~,
\end{equation}
then
\begin{equation}
\sigma_T=\frac{4\pi}{k^2}\sum_\ell (2\ell+1)\IM f_\ell(s)~,
\end{equation}
and
\begin{equation}
\sigma_I=\frac{4\pi}{k^2}\sum_\ell (2\ell+1)\left(\IM f_\ell(s)-|f_\ell(s)|^2\right)~.
\end{equation}
Hence
\begin{equation}
\label{eq:inegi2}
\sigma_I<\frac{4\pi}{k^2}\sum_\ell (2\ell+1)\left(\IM f_\ell-(\IM f_\ell)^2\right)~.
\end{equation}

So while
\begin{align}\label{eq:Imfl}
0&\le \IM f_\ell\le 1~,\\
0&\le \IM f_\ell - (\IM f_\ell)^2\le 1/4~,
\end{align}
However, there is no sharp cut{}-off in the partial wave distribution
and it is not the same distribution which maximizes $\sigma_T$ and $\sigma_I$ for
a given absorptive part: 
\begin{equation}
A_s= \IM F~ , \quad \text{for}\  t<4 m_\pi^2~.
\end{equation}

Here, for simplicity, we shall not use the average given by
(\ref{eq:FBB}), and make the traditional assumption that $A_s$ is a continuous
function of $s$ for fixed $t<4m_\pi^2$. Then, from  (\ref{eq:int1}), we have
\begin{equation}\label{eq:ineqA}
0<A_s(s,t)<\frac{s^2}{\ln s}~,
\end{equation}
on a set of values of $s$ of asymptotic density unity.

We recall  the method to get the bound on $\sigma_T$
total. One tries to maximize
\begin{equation}
\sigma_T \propto\sum_\ell (2\ell+1)\,\IM f_\ell~,
\end{equation}
for a given $A_s$, with $x=1+t/(2 k^2)$
\begin{equation}\label{eq:sixteen}
A_s(s,t)=\sum_\ell (2\ell+1)\,\IM f_\ell\,P_\ell(x)~,
\end{equation}
neglecting the deviation of $\sqrt{s}/(2k)$ from unity.

It is known that the optimal distribution is
\begin{equation}
\begin{aligned}
\IM f_\ell&=1~,\quad \text{for}\ 0\le \ell \le L_T~,\\
\IM f_{L_T+1}&=\epsilon~,\quad 0\le\epsilon\le 1~.
\end{aligned}
\end{equation}
Then we have from (\ref{eq:sixteen})
\begin{equation}
P'_{L_T}(x)+P'_{L_T+1}(x)<\frac{s^2}{\ln s}~,
\end{equation}
Using standard bounds on Legendre polynomials one gets
\begin{equation}
L_T(A_s) \le \frac{k}{\sqrt{t}}\,\ln s~,
\end{equation}
The Froissart bound follows from that:
\begin{equation}
\sigma_T\le \frac{4\pi}{t}\,\ln ^2 s~,
\end{equation}
giving (\ref{eq:FB}) for $ t= 4m_\pi^2$. 
A recent new derivation of this result has been proposed \cite{r11} .

If, on the other hand, we want to maximize $\sigma_I$, where
\begin{equation}
\sigma_I = \sum_\ell (2\ell+1)\left(\IM f_\ell - (\IM f_\ell)^2\right)~,
\end{equation}
we find that the optimal distribution for given $A_s$ is (see \ref{AppA}).
\begin{equation}\label{eq:inegIM}
\IM f _\ell= \frac{1}{2}\left[1-P_\ell(x)/P_{\overline{L}}(x)\right]
\end{equation}
for $0\le \ell\le L_I$, with
\begin{equation}
L_I<\overline{L}<L_I+1~.
\end{equation}
It is obvious that
\begin{equation}
L_I(A_s)> L_T( 2A_s)>L_T (A_s)~.
\end{equation}
Starting from (\ref{eq:inegIM}) one can get a closed expression for $A_s$.
\begin{equation}\label{eq:closedA}
A_s=\frac{1}{2}\left[P'_{L_I}(x)+P'_{L_I+1}(x)-
\frac{(L_I+1)^2 P_{L_I}^2(x)-(x^2-1)P'_{L_I}{}^2}%
{P_{\overline{L}(x)}}
\right]~.
\end{equation}
In fact, we shall not use this expression.
However, since 
\begin{equation}
L_T(s^2/\ln s)\simeq \frac{k}{\sqrt{t}}\,\ln s~,
\end{equation}
$L_I>(k/\sqrt{t}) \ln s$. 
The sum 
\begin{equation}
\frac{1}{2}\sum_0^{L_I} (2\ell+1) 
\left[1-P_\ell\left(1+\frac{t}{2k^2}\right)/P_{\overline{L}}\left(1+\frac{t}{2k^2}\right)\right]\,P_\ell(x)~,
\end{equation}
can split into 
$\sum_{\ell=0}^{L_I-\Delta}+\sum_{\ell= L_I-\Delta}^{L_I}$.
We choose 
\begin{equation}\label{eq:def-delta}
\Delta =\lambda \, k~.
\end{equation}
For $\ell < L_I-\Delta$,
\begin{equation}
\frac{P_\ell(x)}{P_{\overline{L}}(x)}\le \frac{P_\ell(x)}{P_{L_I}(x)}
< \frac{P_{L_I-\Delta}(x)}{P_{L_I}(x)}~.
\end{equation}
We prove, in  \ref{AppB} that
\begin{equation}\label{eq:crude}
\frac{P_{L_I-\Delta}(x)}{P_{L_I}(x)}<4\exp(-\Delta\sqrt{x-1})~,
\end{equation}
(this is a  \emph{very crude bound}, but sufficient for our purpose).

Hence, with the choice (\ref{eq:def-delta}), we get:
\begin{equation}
\frac{1}{2}\sum_{\ell=0}^{L_I-\lambda k}(2\ell+1) \left[1-P_\ell(x)/P_{\overline{L}}(x)\right]
P_\ell(x)>\frac{1}{2}\left[1-4\exp(-\lambda \sqrt{t/2})\right] \sum_{\ell=0}^{L_I-\lambda k} (2\ell+1)P_\ell(x)~.
\end{equation}
So taking  $\lambda=\sqrt{2/t}\,\ln 8$ and $t< 2 k^2$, we get 
\begin{equation}\label{eq:eq31}
\sum_{\ell=0}^{L_I-\lambda k} (2\ell+1)P_\ell(x)<\frac{4\,s^2}{\ln s}~.
\end{equation}
Hence we are back to the same problem as for $\sigma_T$,
except for a change of scale, and we get
\begin{equation}
L_I- \Delta< \frac{k}{\sqrt{t}} \left(\ln s +C\right)~.\end{equation}
Now, from (\ref{eq:def-delta}):
\begin{equation}
L_I <\frac{k}{\sqrt{t}}\left(\ln s+C'\right)~,
\end{equation}
so that
\begin{equation}
\sigma_I<\sum_{\ell=0}^{L_I}(2\ell+1)\,\frac{1}{4}=\frac{k^2\,\ln^2 s}{4\,t}~,
\end{equation}
and
\begin{equation}
\sigma_I< \frac{\pi}{t} \left[\ln^2s+ \mathcal{O}(\ln s)\right]~,
\end{equation}
and, if $t=4 m_\pi^2$
\begin{equation}\label{eq:resu}
\sigma_I<\frac{\pi}{4 m_\pi^2}\,\ln^2 s~.
\end{equation}
There remains of course the fact that (\ref{eq:resu}) holds only on a set
of asymptotic density unity if $A_s$ is a continuous function of $s$ for fixed $t$.
The scale in $s$ cannot be fixed, as it was the case for the total cross section. As we said before, the only  thing we know is that the integral (\ref{eq:int1}) converges for $0\le t < 4m_\pi^2$ sometimes
also for $t=4 m_\pi^2$. For $\sigma_I$,  one would like to have the analogue of
(\ref{eq:FBB}), but, so far, we have not been able to get it.
Another way out is to assume that, beyond a certain energy, $A_s$
is monotonous. The case where it is monotonous decreasing is uninteresting,
and so we take $A_s$ to be monotonous increasing. If
\begin{equation}
I (t ) =\int_{s_0}^\infty \frac{A_s(s,t)}{s^3}\,\mathrm{d} s~,\quad\text{then}\quad
A_s(s,t)< 2 s^2\,I(t)~.
\end{equation}
Then, all constants can be fixed in the bounds on $\sigma_T$ and $\sigma_I$, and the scale problem is removed. Further, if  $I ( t )$ goes to infinity
as $t$ approaches $4m_\pi^2$. We know that $I(t)$ behaves like a negative power of 
$(4m_\pi^2-t)$. By taking $t=4m_\pi^2-1/\ln s$, one can manage to
prove that  (\ref{eq:FB}) and (\ref{eq:eq31}) still hold, with corrective terms of the order of $\ln s \ln(\ln s)$. It is a matter of taste to
decide if this monotonicity assumption is acceptable. Here we shall not give detailed calculations, because we hope to find the analogue of
(\ref{eq:FBB}) for the inelastic cross-section, and to get the best possible
estimates without any artificial assumption. 

This   ends  the rigorous part of this paper. Now comes
the fact that most   theoreticians believe that the worse that can happen at high
energies is that the elastic cross{}-section reaches half of the total cross{}-section, which corresponds to an expanding black disk. This is
the case in the model of Chou and Yang \cite{r12}, and in the model of Cheng and
Wu \cite{r13}, later developed by Bourrely. Soffer and Wu \cite{r14}, and also in
general considerations by Van Hove \cite{r15} who introduces what became known as
the ``overlap function'' which is
\begin{equation}\label{eq:overl}
\sum_\ell (2\ell+1)\left[\IM f_\ell-(\IM f_\ell)^2\right] P_\ell(\cos\theta)~,
\end{equation}
which represents the overlap between inelastic final states produced
by two two{}-body  corresponding to different directions. Here Van hove neglects the real part of the elastic amplitude. From 
\begin{equation} o_\ell=\IM f_\ell-(\IM f_\ell)^2~,\end{equation}
one gets
\begin{equation}
\IM f_\ell=\frac{1\pm \sqrt{1-4\, o_\ell}}{2}~,
\end{equation}

For large $\ell$ one has to choose the \emph{minus} sign, and
Van hove argues that by continuity, or, better analyticity in $\ell$, one has to keep
the minus sign down to $\ell=0$, which means that $\IM f_\ell$ is less than $1/2$.
However, not everybody agrees with this. See for instance the talk of Sergei Troshin in La Londe-les-Maures \cite{r16}. In his view, the scattering
amplitude becomes dominantly elastic in the high energy limit. To say the least, this seems to me extremely unlikely and, therefore, I tend to believe
that we have
\begin{equation}\label{eq:FB-half}
\sigma_T<\frac{1}{2}\,\frac{\pi}{m_\pi^2} \,\ln^2 s~,
\end{equation}

Certainly, this is not enough to satisfy Peter Landshoof but it represents nevertheless a progress. 
\subsection*{Acknowledgments}
I would like to thank warmly Jacques Soffer for his invitation
to the workshop of La Londe-les-Maures in September 2008. It is in this stimulating atmosphere that I made this little step forward. I would
also like to thank David Atkinson, Porter Johnson, Peter Landshoff,
Jean-Marc Richard, Shasanka Roy, and Tai Tsun Wu for stimulating discussions.

\appendix
\let\thesec=\thesection
\renewcommand{\thesection}{Appendix\ \thesec}
\renewcommand{\theequation}{\thesec.\arabic{equation}}
\section{\ \ }\label{AppA}
\setcounter{equation}{0}
Calling $\IM f_\ell = y_\ell$,we try to maximize
\begin{equation}
\sigma_I=\sum_\ell (2\ell+1)\,(y_\ell-y_\ell^2)~,
\end{equation}
for given
\begin{equation}
A_s=\sum_\ell (2\ell+1)\,y_\ell\,P_\ell(x)~,\quad x>1~.
\end{equation}
We start with a heuristic variationnal argument. We have
\begin{align}
\delta A_s=0&=\sum_\ell (2\ell+1)\,\delta y_\ell\, P_\ell~,\\
\delta \sigma_I=0&=\sum_\ell (2\ell+1)\,\delta y_\ell\, (1-2 y_\ell)~.
\end{align}
Hence , using a Lagrange multiplyer:
\begin{equation}y_\ell=\frac{1}{2}\,[1-c\,P_\ell(x)]~.
\end{equation}
This is, in fact the correct answer. We shall prove it.

Assume that $\{y_\ell\}$ is the maximizing distribution.
Consider only two terms, $y_\ell$ and $y_L$. another distribution contains
$y_\ell+\Delta y_\ell$ and $y_L+\Delta y_L$. $A_s$ is fixed.
Hence
\begin{equation}\label{A6}
(2\ell+1)\, \Delta_\ell\, P_\ell+(2L+1) \Delta_L\,P_L=0~.
\end{equation}
On the other hand
\begin{equation}
\Delta \sigma_I=(2\ell+1)\Delta_\ell(1-2y_\ell)+(2L+1)\Delta_L(1-2y_L)-(2\ell+1)\Delta_\ell^2-(2L+1)\Delta_L^2~.
\end{equation}
If we choose 
\begin{equation}\label{A8}
1-2\,y_\ell= c\,P_\ell~,\qquad 1-2\,y_L= c\,P_L~,
\end{equation}
we get, from (\ref{A6})
\begin{equation}
\Delta \sigma_I=-(2\ell+1)\Delta_\ell^2-(2L+1)\Delta_L^2<0~,
\end{equation}
Hence the choice (\ref{A8}) maximizes $\sigma_I$.
Therefore, we take
\begin{equation}
y_\ell=\frac{1}{2}\left[ 1 -c \,P_\ell(x)\right]~.
\end{equation}
Now, what is $c$?  If the sum is
\begin{equation}
A_s=\frac{1}{2}\sum_{\ell=0}^{\ell=L_I} (2\ell+1)\,P_\ell(x)\,[1-c\,P_\ell(x)]~,
\end{equation}
obviously  $0<c<1/P_{L_I}$
because $0<\IM f_\ell <1$. But it is not possible for $c$ to be less than
$1/P_{L_I+1}$, because we could apply our previous reasoning to the last two partial waves, the
last one being zero. This would lead to changing $c$.  So
\begin{equation}
\frac{1}{P_{L_I}}<c<\frac{1}{P_{L_I+1}}~.
\end{equation}
Now we give, for completeness , in the case where $c= 1/P_{L_I}$
exactly, the complete expression for $A_s$, even though we don't
use it.  From Gradshtein and Ryzhik \cite{r17}, we get
\begin{equation}
\sum_{\ell=0}^L (2\ell+1)\,P_\ell^2=(L+1)\left(P'_{L+1}\,P_L-P'_L\,P_{L+1}\right)~,
\end{equation}
and so
\begin{equation}
A_s=\frac{1}{2}\left[P'_L+P'_{L+1}-(L+1)\left(P'_{L+1}-P'_L\,P_{L+1}/P_{L}\right)\right]~.
\end{equation}
Notice that $A_s$ vanishes for $x=1$. It is possible to get an expression 
with $x-1$ explicitly factored out, using the Legendre differential 
equation and recursive relations.

\section{\ \ }\label{AppB}
\setcounter{equation}{0}
We derive a upper bound on%
\begin{equation}
P_{L-\Delta}(x)/P_L(x)~,\quad x>1~.
\end{equation}
From
\begin{equation}\label{B2}
P_\ell=\frac{1}{\pi}\int_0^\pi \left(x+\sqrt{x^2-1}\,\cos \phi\right)^\ell\,\mathrm{d}\phi~,
\end{equation}
we get, using the Minkowsky--H{\"o}lder inequality, for $x>1$,
\begin{equation}
P_{L-\Delta}(x) <[ P_L(x)]^{(L-\Delta)/L}~,
\end{equation}
so
\begin{equation}
\frac{P_{L-\Delta}(x)}{P_L(x)}<\frac{1}{[ P_L(x)]^{\Delta/L}}~.
\end{equation}
Now, we need a lower bound for $P_\ell$ . A very crude lower bound is enough: cutting the integral (\ref{B2}) at $\phi=\pi/4$ , we get
\begin{equation}
P_\ell(x)>\frac{1}{4}\left[x+\sqrt{\frac{x^2-1}{2}}\right]^\ell~,
\end{equation}
and, for $1<x<7$,
\begin{equation}\label{B5}
P_\ell(x)>\frac{1}{4}\,\exp(\ell\sqrt{x-1})~.
\end{equation}
Since $x=1+t/(2 k^2)$ and $t<4m_\pi^2$,
this corresponds to $k>0.4\,m_\pi$, ridiculously small in
these high energy considerations.

One could do much better than that. For instance, S.M. Roy \cite{r18}
quotes an unpublished optimal result of mine
\begin{equation}
P_\ell(x)> \frac{2N!}{(N!)^2}\left(\frac{N+1}{2(2N+1)}\right)^N\left(x+\frac{N}{N+1}\sqrt{x^2-1}\right)^\ell~,
\end{equation}
for $N = 1, 2, 3, 4\ldots$ For instance
\begin{equation}
P_\ell(x)> \frac{54}{100}\left(x+\frac{2}{3}\sqrt{x^2-1}\right)^\ell~.
\end{equation}

If we take $\ell=2$ and $\ell=3$, we see that this bound is saturated
for $x\to\infty$. However, these refinements are not really needed for our purpose. Inequality (\ref{B5}) is enough.

\end{document}